# Angular dependence of domain wall resistivity in artificial magnetic domain structures

A. Aziz[1], S. J. Bending[1], H. G. Roberts[1], S. Crampin[1], P. J. Heard[2] and C. H. Marrows[3]

[1]Department of Physics, University of Bath, Claverton Down, Bath BA2 7AY, UK

[2]Interface Analysis Centre, University of Bristol, Bristol BS2 8BS, UK

[3]School of Physics and Astronomy, University of Leeds, Leeds LS2 9JT, UK

We exploit the ability to precisely control the magnetic domain structure of perpendicularly magnetized Pt/Co/Pt trilayers to fabricate artificial domain wall arrays and study their transport properties. The scaling behaviour of this model system confirms the intrinsic domain wall origin of the magnetoresistance, and systematic studies using domains patterned at various angles to the current flow are excellently described by an angular-dependent resistivity tensor containing perpendicular and parallel domain wall resistivities. We find that the latter are fully consistent with Levy-Zhang theory, which allows us to estimate the ratio of minority to majority spin carrier resistivities, $\rho_\downarrow/\rho_\uparrow \sim 5.5$, in good agreement with thin film band structure calculations.

72.15.Gd, 75.60.Ch , 75.70.Kw, 73.40.-c

The enormous potential of devices based upon the simultaneous manipulation of electronic charge and spin – *spintronics* – has led the study of domain wall physics to be pursued with renewed vigor. Transport across domain walls in magnetic semiconductors [1], oxides [2] and metals [3] has been investigated, including studies of constrictions where spin reversal is more abrupt and magnetoresistance can be dramatically increased [4]. As well as their fundamental importance to an understanding of the physics of magnetoresistive devices [5], domain walls are also central to novel logic devices [6] and, through the phenomenon of spin transfer torque domain wall motion [7], provide a strong contender for future data storage cells.

Typical values of the *intrinsic* domain wall magnetoresistance in ferromagnetic thin films are only of the order of 0.1-1.0 percent of the normal resistivity, and are readily masked by *extrinsic* effects associated with the magnetic domain structure [8]. This has made it difficult to distinguish between different theoretical models of the resistance which variously predict *positive* domain wall resistance, due to spin mistracking in the domain wall [9], *negative*, due to the suppression of weak localisation corrections [10], or *of either sign*, depending on the spin-dependent relaxation times of the charge carriers [11]. By considering spin-dependent impurity scattering within the wall, akin to the theory of giant magnetoresistance in magnetic multilayers, Levy and Zhang (LZ) [12] have developed a theory of *positive* domain wall resistance for current flow both perpendicular to (CPW) and in the plane of (CIW) the wall, which has been used to interpret a number of experimental results in thin ferromagnetic films [13,14,15]. However, problems associated both with the elimination of extrinsic effects and in making quantitative estimates of the physical quantities that enter the LZ theory, have made it difficult to claim conclusive agreement.

Here we report results on a controllable 'artificial domain' system which allow us entirely to exclude extrinsic effects, as well as to study precisely the systematic dependence of domain wall resistance on the angle between the direction of current and the domain wall. Comparison with a numerical simulation of our sample structures reveals that the full wall resistance is only measured for tilted structures with a sufficiently large domain width to wire width aspect ratio. Measurements on suitably designed samples permit us, for the first time, to map precisely the angular dependence of the domain wall resistivity tensor. These results are fully consistent with Levy-Zhang theory, which allows us to make a quantitative estimate of the ratio of minority to majority spin carrier resistivities, $\rho_\downarrow/\rho_\uparrow$, that is in good agreement with thin film band structure calculations.

Our results are based upon the magnetotransport properties of *artificial* magnetic domain structures in perpendicularly magnetised Pt/Co/Pt trilayers, which are realised by locally modifying the anisotropy by irradiation with a Ga focused ion beam (Ga-FIB) [16]. By depositing thin $SiO_2$ capping layers prior to Ga-FIB irradiation (which is known to cause mixing/roughening at the Co/Pt interfaces and a relief of interface strain) we have shown that we can precisely control the coercive fields of the artificial domains [17]. *In this way magnetic domain structures can be defined of arbitrary shape with <10nm spatial resolution, whose domain walls lie at precisely controlled angles,* as exemplified in Figs. 1a,b. Other attractive features of this system include the fact that closure domains, which can be a major source of anisotropic magnetoresistance [18], are not energetically favoured in ultrathin (0.6 nm) Co films with strong perpendicular anisotropy, whilst the relatively high film resistivity (~21 μΩ cm at 300 K) leads to short scattering times, $\tau$, and $\omega_c\tau \sim 10^{-4} \ll 1$ (where $\omega_c$ is the cyclotron frequency), and allow Lorentz magnetoresistance and other effects associated with the wiggling of current lines at domain walls to be completely neglected.

Our transport devices were fabricated from a Pt(3.5 nm)/Co(0.6 nm)/Pt(1.6 nm) sandwich structure, which was deposited on a Si/$SiO_2$ substrate at 300 K using dc magnetron sputtering. Optical lithography and reactive ion etching were used to pattern the magnetic films in a Wheatstone bridge geometry based on 1μm wide 'active' wires (Fig. 1c). After deposition of an additional 8 nm SiO$_2$ capping layer, the desired domain pattern was irradiated with a 30 keV Ga ion dose of 0.009pC/μm$^2$ in two asymmetric arms of the bridge (Fig. 1d) using an FEI Strata 201 Ga-FIB. To eliminate any extraordinary Hall effect originating from the Pt/Co/Pt underneath the voltage contacts, the Pt capping layer and Co layer were etched away in an Ar plasma after photolithography, and before evaporation of Ti/Au electrodes. It is possible that a very thin film of Co remained unetched after this step. The magnetic force microscopy (MFM) images shown in Figs. 1a,b demonstrate the precision with which 'striped' artificial domains can be defined. Images a1 and b1 were captured at $H_\perp$ = 50 Oe when the magnetisation of the irradiated regions was reversed with respect to unirradiated regions, while a2 and b2 show the same regions after the magnetisation has been saturated everywhere at $H_\perp$ = 300 Oe. The very weak contrast in these latter images indicates that Ga-FIB irradiation causes a very small change in the saturation magnetisation due to Co/Pt mixing, the resistivity of these regions also increases very slightly by ~2.5%. Measurements of our bridge devices were performed with a 10μA ac current (29 Hz) between current contacts A and B, and the output voltage measured with a lock-in amplifier between V1 and V2. An external magnetic field was applied perpendicular to the plane of the device with a solenoid.

Fig. 2a illustrates a set of magnetoresistance measurements recorded at 300K on bridges with stripe domain 'superlattice' structures containing 0, 2, 4 or 8 artificial domains (all 500nm wide in CPW geometry). Each measurement has been performed on a different, but otherwise identical, bridge structure fabricated on the same chip. For H > +150 Oe, both the irradiated and unirradiated domains are magnetised in the same direction and the resistance is at a low value. If the field is now reversed to H ≈ -50 Oe, the magnetisation of the irradiated stripe domains, which have a smaller coercive field [17], begin to reverse and the resistance increases to a maximum when domain walls are fully developed between irradiated and unirradiated regions. When the field is made more negative still it eventually reaches the coercive field of the unirradiated stripes (~100 Oe), which in turn start to reverse and the resistance drops back down to a minimum. The same behaviour is observed if the field is now swept back up in a positive sense. The weakly varying background signal is probably due to magnetoresistance arising from residual Co left under the voltage contacts after etching as described above. A separate Hall bar device was patterned in the same Pt/Co/Pt trilayer structure with Hall 'crosses' containing unirradiated and irradiated (with the same Ga-FIB dose) strips of the same width for extraordinary Hall effect (EHE) investigations [17]. The coercive fields inferred from these EHE measurements agree well with the values estimated from the magnetoresistance data. Fig. 2b shows that the magnetoresistance of such domain structures increases linearly with the number of domain walls. Also included are data for structures with different irradiated domain widths, which confirm that the resistance is independent of the width of artificial domains, exactly as expected for intrinsic domain wall magnetoresistance at the border between the irradiated and unirradiated regions.

Having established the *intrinsic* domain wall origin of the magnetoresistance, magnetic domains were patterned at various angles with respect to the current flow in order to probe the perpendicular and parallel domain wall resistivities. Multiple artificial magnetic domains were patterned, as described above, with the normal to the wall subtending an angle $\theta$ in the range 0°-60° with respect to the current direction. The geometry is shown in the inset of Fig. 3b. For all samples the channel width was 1μm, and the width of the irradiated artificial domains were kept invariant at 1μm. The 16μm long Pt/Co/Pt wires imposed geometrical constraints on the number of superlattice repeats that could be used. For $\theta$=0°, 20°, and 30° six irradiated stripes were patterned, whereas for $\theta$=40°, 50° and 60° only three irradiated stripes were possible. Therefore the magnetoresistance is due to twelve domain walls for the former set of angles, and six for the latter. Fig. 3a illustrates the 300K magnetoresistance of our bridge devices with different tilt angles. A steady decrease in the maximum resistance change is observed as the angle, $\theta$, is increased. Also noticeable is a systematic broadening of the resistance peaks at larger angles. Since the domain width is the same in each case this can only arise from changes in the reversal dynamics due to angle-dependent changes in the magnetostatic interaction at the sharp and blunt corners of tilted domains. This will, for example, lead to a distribution of de-pinning fields for a domain wall trapped at the interface between an irradiated and an unirradiated domain.

To understand the angular dependence of the magnetoresistance, we model the current flow within the device channel, treating the problem as two-dimensional. In steady state, the continuity equation requires that div$J$=0, and from Maxwell's equations curl$E$=0, where the current density $J$ and electric field $E$ are related via the resistivity tensor $E=\rho J$. On account of the high resistivity of our transport devices we neglect off-diagonal terms of $\rho$ within each region, which is either (i) unirradiated, with $\rho_{xx}=\rho_{yy}=\rho_0$ (we take $x$ to be along the device channel), (ii) irradiated, with $\rho_{xx}=\rho_{yy}=\rho_1$, or (iii) a domain wall, for which

$$\rho_{DW} = \begin{pmatrix} \cos\theta & \sin\theta \\ -\sin\theta & \cos\theta \end{pmatrix} \begin{pmatrix} \rho_{CPW} & 0 \\ 0 & \rho_{CIW} \end{pmatrix} \begin{pmatrix} \cos\theta & -\sin\theta \\ \sin\theta & \cos\theta \end{pmatrix}, \qquad (1)$$

with $\rho_{CPW}$ and $\rho_{CIW}$ the resistivities for current flow perpendicular and parallel to the domain wall respectively [12]. We solve the equations numerically in terms of a stream function, $\psi$, with $J=(\partial\psi/\partial y,-\partial\psi/\partial x)$, assuming uniform current $J=J_0=(J_0,0)$ at each end of the device channel and no current flow through the sides. The equations are discretised on a piecewise uniform finite difference grid, and solved iteratively using multi-grid relaxation with continuity of the normal current and tangential electric field boundary conditions at each interface. Iterations are terminated when the residual is less that $10^{-10}$, and the potential $V$ along the device channel then calculated from $\mathbf{E}=-\text{grad}V$.

In Fig. 3b we show typical results for the change in the potential along the device channel created by the presence of domain walls, for different angles of the irradiated region. As well as a sharp increase across the domain wall itself, the potential also displays an increase on both sides, which is due to the formation of static eddy-currents, shown in the inset. The size and range of this contribution increases with the angle $\theta$, but our simulations confirm that for the structures investigated here the potential change is fully contained within the device channel and not responsible for the changes in Fig. 3a. To understand the longitudinal channel resistance, we note that the *additional* resistivity due to the domain wall, $\delta\rho_{CPW}=\rho_{CPW}-\rho_0$, $\delta\rho_{CIW}=\rho_{CIW}-\rho_0$, is small. Neglecting the difference in the resistivity of the irradiated and unirradiated regions (which changes the final result by at most a few percent) the electric field can be written as $\mathbf{E}=(\boldsymbol{\rho}_0+\delta\boldsymbol{\rho})(\mathbf{J}_0+\delta\mathbf{J})$ so that to first order $\delta E_x=\delta\rho_{xx}J_0+\rho_0\delta J_x$. Since $\delta J_x$ integrates to zero along the channel, the additional *resistance* may be attributed to $\delta\rho_{xx}$ and hence reflects the angular variation of the resistivity tensor in Eqn. (1), which in general contains both CPW and CIW contributions:

$$\delta\rho_{DW} = \delta\rho_{xx} = (\delta\rho_{CPW} - \delta\rho_{CIW})\cos^2\theta + \delta\rho_{CIW} \qquad (2)$$

In Fig. 3c we plot the domain wall resistivity, derived from curves of the type shown in Fig. 3a, against $\cos^2\theta$. Also plotted are the results of our numerical simulations. Both measured and simulated results are accounted for very well by Eq. (2). In order to derive $\delta\rho_{DW}$ from our measurements we have taken the domain wall width in our 0.6nm thick Co layer to be 15 nm [18], and assume a parallel resistor model for our trilayer films with resistivities uniformly scaled from the bulk values to give the correct total resistance.

Unlike previous attempts to study the anisotropy of domain wall resistivity [2,14], which used naturally-occurring reasonably well-ordered striped domain structures, these are the first measurements to exploit precise artificial control of domain wall orientation. This eliminates much of the uncertainty involved in analysis of the observed angular dependence. For example, we do not need to estimate the relative proportion of domain walls aligned perpendicular to the current [14]. As expected from Eq. (2) and the theory of Levy and Zhang, which predicts $\delta\rho_{CPW} > \delta\rho_{CIW}$, we find that the domain wall resistivity remains positive for all angles, with values for the perpendicular and parallel domain wall resistivities that may be inferred from our measurements of $\delta\rho_{CPW} = 23.1\pm1.1\times10^{-3}$ µΩ-cm and $\delta\rho_{CIW} = 3.5\pm0.7\times10^{-3}$ µΩ-cm respectively. We note that the results in Fig. 3c also limit the magnitude of any non-Hall-like contribution to the domain wall resistivity tensor – which has been assumed to be zero in Eq. (1) but which, if present, would make a $\sin 2\theta$ contribution – to less than $0.5\times10^{-3}$ µΩ-cm. The magnetoresistivity ratio $MR_{CPW} = \delta\rho_{CPW}/\rho_0$ is 0.1%, which is comparable to values deduced for domain walls in other metallic systems [3]. The ratio of CPW to CIW resistivities is $\delta\rho_{CPW}/\delta\rho_{CIW} \approx 6.6\pm1.2$; according to LZ theory this is $3+10\sqrt{\rho_\downarrow/\rho_\uparrow}/(\rho_\downarrow/\rho_\uparrow +1)$ where $\rho_\downarrow/\rho_\uparrow$ is the ratio of minority to majority spin carrier resistivities. The value $\rho_\downarrow/\rho_\uparrow\sim5.5$ which we deduce is greater than in bulk Co ($\rho_\downarrow/\rho_\uparrow\sim4$) [19], but consistent with the Fermi level density of states ratio that we find in band structure calculations of a Pt/3ML Co/Pt sandwich structure, where an enhancement of the minority-spin density of states arises due to band-narrowing [20].

This model system provides exciting opportunities for both studying and exploiting domain wall physics, such as current-induced domain wall motion. Our accomplishments in controlling the exact domain structure that forms in the nanowire and suppressing extrinsic effects are important ones, eliminating difficulties that have plagued the domain wall resistance field for many years. These measurements have demonstrated that an unambiguous smooth mapping of the DW resistivity from CPW to CIW can be observed in carefully designed samples containing tilted domains. A natural development will be to study whether the magnitude of magnetoresistive effects can be enhanced by fabricating nanoconstrictions within the nanowire, or whether there are important effects arising from reducing the wire width/domain width ratio for tilted domains. Multiple-terminal "mesoscopic" measurements to probe transverse resistance *within* tilted domains will provide further insight into the validity of microscopic theories of domain wall scattering. Finally, our room temperature measurements point to applications in practical devices – one can envisage, for example, using the domain wall angle as a control parameter in multiple-state logic devices based upon spin-torque transfer.

This work was supported by the Leverhulme Trust, through grant no. F/00 351/F.

**References**

[1]  e.g. H. X. Tang *et al*, Nature **431**, 52 (2004); D. Chiba *et al*, Phys. Rev. Lett. **96**, 096602 (2006).

[2]  L. Klein, *et al.*, Phys. Rev. Lett. **84**, 6090 (2000); M. Feigenson, L. Klein, J. W. Reiner, and M. R. Beasley, Phys. Rev. B **67**, 134436 (2003).

[3]  For recent reviews see: C.H. Marrows, Adv. in Phys. **54**, 585 (2005); A. D. Kent, J. Yu, U. Rudiger, and S. S. P. Parkin, J. Phys.: Condens. Matter **13**, R461 (2001).

[4]  P. Bruno, Phys. Rev. Lett. **83**, 2425 (1999); C. Ruster *et al.*, Phys. Rev. Lett. **91**, 216602 (2003); S. Lepadatu and Y.B. Xu, Phys. Rev. Lett. **92**, 127201 (2004).

[5]  S. S. P. Parkin, Phys. Rev. Lett. **71**, 1641 (1993).

[6]  D.A. Allwood *et al.*, Science **296**, 2003 (2002); D.A. Allwood *et al.*, Science **309**, 1688 (2005).

[7]  G. Tatara and H. Kohno, Phys. Rev. Lett. **92**, 086601 (2004); S. Zhang and Z. Li, Phys. Rev. Lett. **93**, 127204 (2004).

[8]  I. Knittel and U. Hartmann, J. Magn. Magn. Mater. **294**, 16 (2005); U. Rudiger, J.Yu, L. Thomas, S. S. P. Parkin, and A. D. Kent, Phys. Rev. B **59**, 11914 (1999).

[9]  M. Viret, *et al.*, Phys. Rev. B **53**, 8464 (1996).

[10]  G. Tatara and H. Fukuyama, Phys. Rev. Lett. **78**, 3773 (1997).

[11]  R. P. van Gorkom, A. Brataas, and G. E. W. Bauer, Phys. Rev. Lett. **83**, 4401 (1999).

[12]  P. M. Levy and S. Zhang, Phys. Rev. Lett. **79**, 5110 (1997).

[13]  D. Ravelosona, *et al.*, Phys. Rev. B **59**, 4322 (1999).

[14]  M. Viret, *et al.*, Phys. Rev. Lett. **85**, 3962 (2000).

[15]  C.H. Marrows and B.C. Dalton, Phys. Rev. Lett. **92**, 097206 (2004).

[16]  T. Devolder, Phys. Rev. B **62**, 5794 (2000).

[17]  A. Aziz, S. J. Bending, H. G. Roberts, S. Crampin, P. J. Heard, and C. H. Marrows, J. Appl. Phys. **98**, 124102 (2005); *ibid* **99**, 08C504 (2006).

[18]  I. Knittel and U. Hartmann, J. Magn. Magn. Mater. **294**, 16 (2005); U. Rudiger, J. Yu, L. Thomas, S. S. P. Parkin, and A. D. Kent, Phys. Rev. B **59**, 11914 (1999).

[19]  D. A. Papaconstantopoulos, *Handbook of The Band Structure of Elemental Solids*, (Plenum Press, New York, 1986).

[20]  H. G. Roberts, PhD Thesis, University of Bath (2006) (unpublished).

**Figure Captions**

FIG. 1 (a1) and (b1) show MFM images of 90° and 45° domain 'superlattices' created by Ga-FIB irradiation; (a2) and (b2) show MFM images of the same regions after saturation of the magnetisation at 300 Oe. (c) Micrograph of the Wheatstone bridge device consisting of four Pt/Co/Pt wires each 16μm long and 1μm wide. (d) Schematic of the top pair of Pt/Co/Pt wires of the Wheatstone bridge.

FIG. 2 (a) Magnetoresistance (offset vertically for clarity) of bridge structures containing 0, 2, 4 and 8 irradiated strips in CPW geometry at 300K. (b) Maximum resistance change as a function of the number of irradiated strips of various widths, d.

FIG. 3 (a) RT Magnetoresistance (offset vertically for clarity) of bridge structures containing irradiated strips patterned at different angles relative to the current flow. (b) Calculated potential profile along the middle of the device channel containing a single patterned strip 1μm wide at different angles (lower inset). Also shown are typical static eddy currents induced by the domain wall (DW) – a 30nm region containing the DW has been expanded horizontally to enable the current lines there to be resolved. (c) Angular dependence of the total domain wall resistivity; diamonds: measured values from (a); circles: calculations such as in (b), using $\delta\rho_{CPW}$=23 ×$10^{-3}$ μΩ-cm, $\delta\rho_{CIW}$=3.5 ×$10^{-3}$ μΩ-cm, DW width 15nm; line: Eq. (2) using the same values.

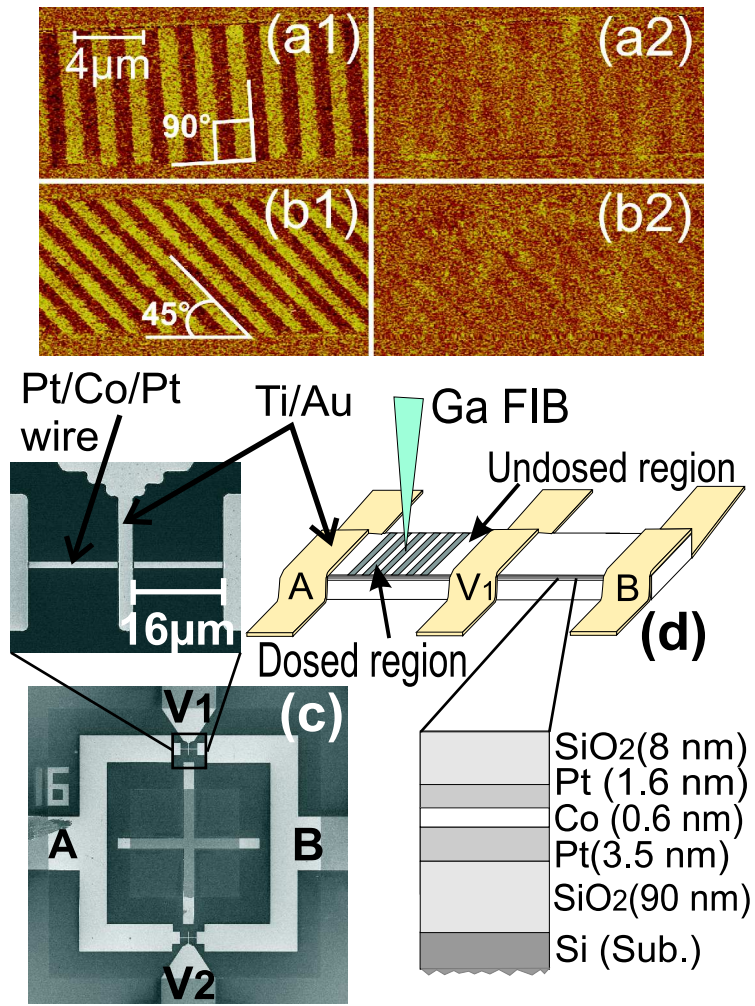

Fig. 1 *"Angular dependence of domain wall resistivity...",* A.Aziz *et al.*

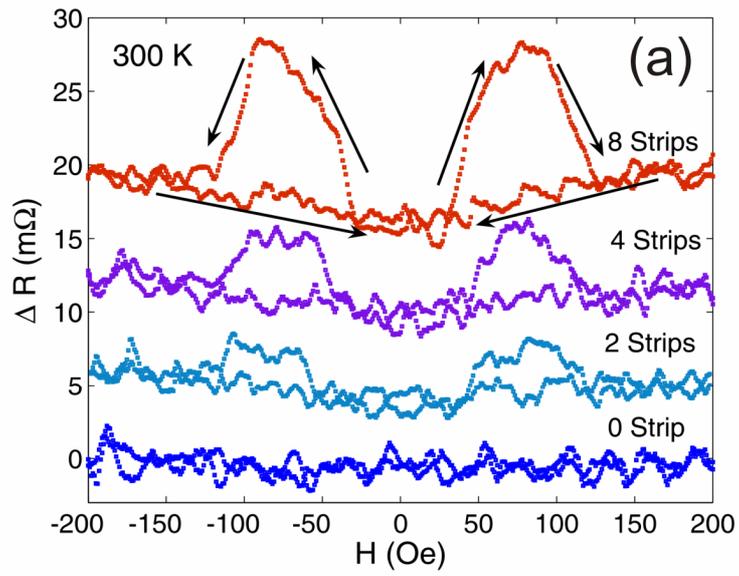

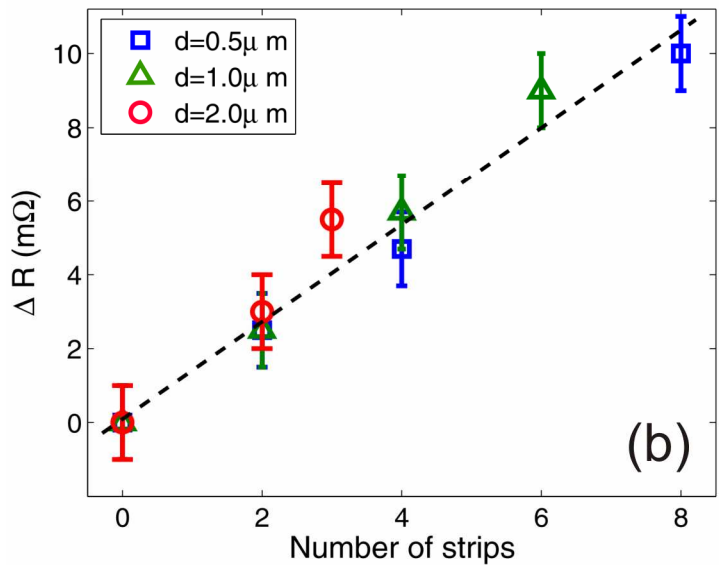

Fig. 2 "*Angular dependence of domain wall resistivity*...", A.Aziz *et al*.

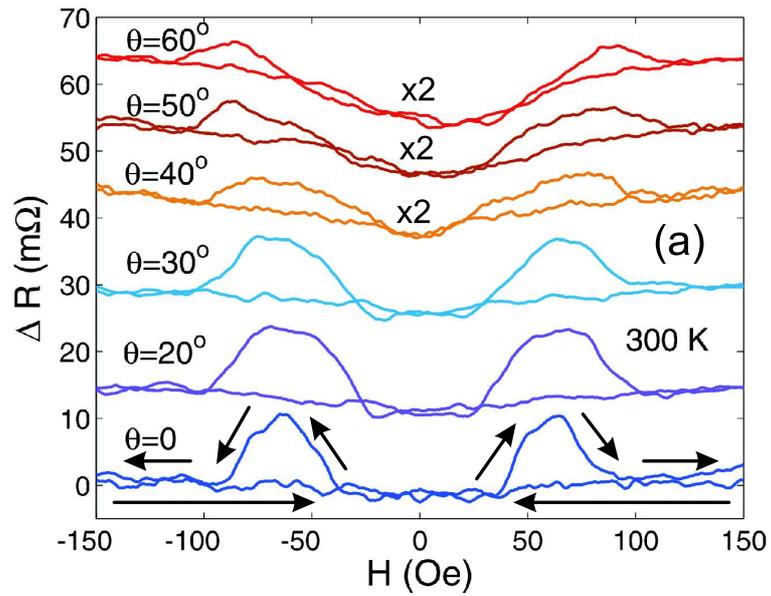

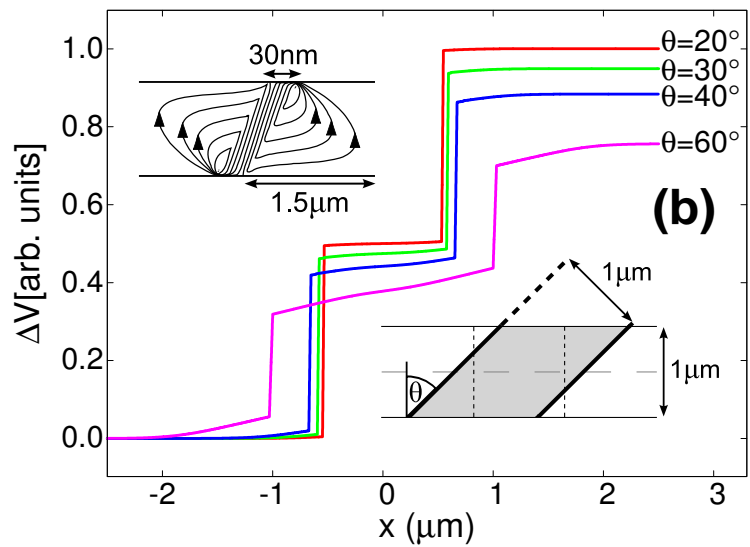

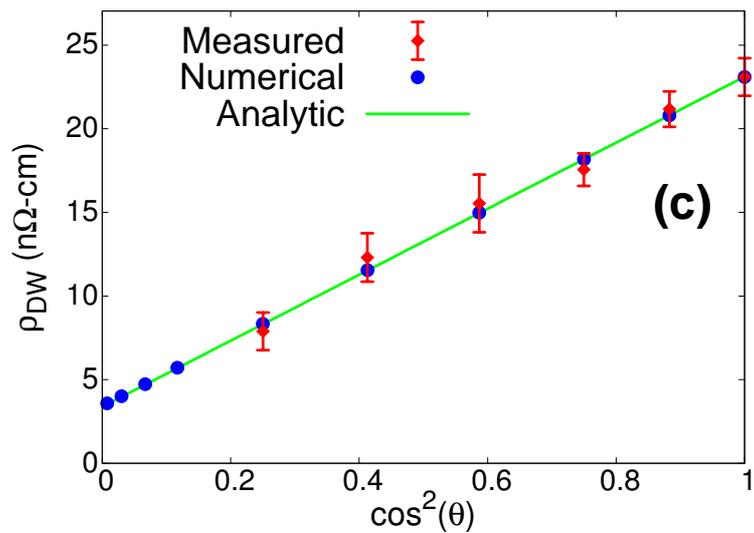

14